\documentclass[%
 reprint,
 superscriptaddress,
 amsmath,amssymb,
 aps,
  aps,                
  prd
]{revtex4-2}

\usepackage{hyperref} 
\usepackage{orcidlink}
\usepackage{graphicx}
\graphicspath{{figs/}}
\usepackage{dcolumn}
\usepackage{bm}
\usepackage{xcolor}
\usepackage{hyperref}
\usepackage{amsmath}
\usepackage[mathlines]{lineno}

\newcommand{\polarbear}{\textsc{POLARBEAR}}

\newcommand{\be}{\begin{equation}}
\newcommand{\ee}{\end{equation}}
\newcommand{\bea}{\begin{eqnarray}}
\newcommand{\eea}{\end{eqnarray}}

\newcommand\acb{\ensuremath{A_{\rm CB}}}
\newcommand\bpc{\ensuremath{B_{\rm 1\,Mpc}}}

\newcommand*\Bell{\ensuremath{\boldsymbol\ell}}

\newcommand{\lcdm}{$\Lambda$CDM}

\newcommand{\upperscaleinvar}{\ensuremath{A_{\textrm{CB}}<0.027}}
\newcommand{\upperscale}{\ensuremath{A_{\textrm{CB}}<0.040}}

\newcommand{\upperscaleBB}{\ensuremath{A_{\textrm{CB}}<0.53}}

\newcommand{\upperscaleinvarsptpol}{\ensuremath{A_{\textrm{CB}}<0.1 }}
\newcommand{\upperscaleBinvar}{\ensuremath{B_{\rm 1\,Mpc} < 9.6\,{\rm nG}}}

\newcommand{\upperscaleginvar}{\ensuremath{g_{\alpha\gamma} < 1.7\times 10^{-2}/ H_I}}

\newcommand{\planck}{{\it Planck}}

\newcommand{\sqdeg}{\mbox{deg$^2$}}

\begin{document}

\preprint{APS/123-QED}

\title{Searching for Anisotropic Cosmic Birefringence with SPT-3G}


\author{A.~Coerver\,\orcidlink{0000-0002-2707-1672}}
\email{acoerver@berkeley.edu}
\affiliation{Department of Physics, University of California, Berkeley, CA, 94720, USA}
\author{M.~Rahimi}
\affiliation{School of Physics, University of Melbourne, Parkville, VIC 3010, Australia}
\author{D.~Anbajagane}
\affiliation{Department of Astronomy and Astrophysics, University of Chicago, 5640 South Ellis Avenue, Chicago, IL, 60637, USA}
\affiliation{Kavli Institute for Cosmological Physics, University of Chicago, 5640 South Ellis Avenue, Chicago, IL, 60637, USA}
\author{E.~Anderes\,\orcidlink{0009-0003-3245-3979}}
\affiliation{Department of Statistics, University of California, One Shields Avenue, Davis, CA 95616, USA}
\author{A.~J.~Anderson\,\orcidlink{0000-0002-4435-4623}}
\affiliation{Department of Astronomy and Astrophysics, University of Chicago, 5640 South Ellis Avenue, Chicago, IL, 60637, USA}
\affiliation{Kavli Institute for Cosmological Physics, University of Chicago, 5640 South Ellis Avenue, Chicago, IL, 60637, USA}
\affiliation{Fermi National Accelerator Laboratory, MS209, P.O. Box 500, Batavia, IL, 60510, USA}
\author{B.~Ansarinejad}
\affiliation{School of Physics, University of Melbourne, Parkville, VIC 3010, Australia}
\author{M.~Archipley\,\orcidlink{0000-0002-0517-9842}}
\affiliation{Department of Astronomy and Astrophysics, University of Chicago, 5640 South Ellis Avenue, Chicago, IL, 60637, USA}
\affiliation{Kavli Institute for Cosmological Physics, University of Chicago, 5640 South Ellis Avenue, Chicago, IL, 60637, USA}
\author{L.~Balkenhol\,\orcidlink{0000-0001-6899-1873}}
\affiliation{Sorbonne Universit\'e, CNRS, UMR 7095, Institut d'Astrophysique de Paris, 98 bis bd Arago, 75014 Paris, France}
\author{D.~R.~Barron\,\orcidlink{0000-0002-1623-5651}}
\affiliation{Department of Physics and Astronomy, University of New Mexico, Albuquerque, NM, 87131, USA}
\author{P.~S.~Barry\,\orcidlink{0000-0001-9103-9354}}
\affiliation{School of Physics and Astronomy, Cardiff University, Cardiff, CF24 3AA, UK}
\author{K.~Benabed}
\affiliation{Sorbonne Universit\'e, CNRS, UMR 7095, Institut d'Astrophysique de Paris, 98 bis bd Arago, 75014 Paris, France}
\author{A.~N.~Bender\,\orcidlink{0000-0001-5868-0748}}
\affiliation{Department of Astronomy and Astrophysics, University of Chicago, 5640 South Ellis Avenue, Chicago, IL, 60637, USA}
\affiliation{Kavli Institute for Cosmological Physics, University of Chicago, 5640 South Ellis Avenue, Chicago, IL, 60637, USA}
\affiliation{High-Energy Physics Division, Argonne National Laboratory, 9700 South Cass Avenue, Lemont, IL, 60439, USA}
\author{B.~A.~Benson\,\orcidlink{0000-0002-5108-6823}}
\affiliation{Department of Astronomy and Astrophysics, University of Chicago, 5640 South Ellis Avenue, Chicago, IL, 60637, USA}
\affiliation{Kavli Institute for Cosmological Physics, University of Chicago, 5640 South Ellis Avenue, Chicago, IL, 60637, USA}
\affiliation{Fermi National Accelerator Laboratory, MS209, P.O. Box 500, Batavia, IL, 60510, USA}
\author{F.~Bianchini\,\orcidlink{0000-0003-4847-3483}}
\affiliation{Kavli Institute for Particle Astrophysics and Cosmology, Stanford University, 452 Lomita Mall, Stanford, CA, 94305, USA}
\affiliation{Department of Physics, Stanford University, 382 Via Pueblo Mall, Stanford, CA, 94305, USA}
\affiliation{SLAC National Accelerator Laboratory, 2575 Sand Hill Road, Menlo Park, CA, 94025, USA}
\author{L.~E.~Bleem\,\orcidlink{0000-0001-7665-5079}}
\affiliation{Department of Astronomy and Astrophysics, University of Chicago, 5640 South Ellis Avenue, Chicago, IL, 60637, USA}
\affiliation{Kavli Institute for Cosmological Physics, University of Chicago, 5640 South Ellis Avenue, Chicago, IL, 60637, USA}
\affiliation{High-Energy Physics Division, Argonne National Laboratory, 9700 South Cass Avenue, Lemont, IL, 60439, USA}
\author{S.~Bocquet\,\orcidlink{0000-0002-4900-805X}}
\affiliation{University Observatory, Faculty of Physics, LMU Munich, Scheinerstr.~1, 81679 Munich, Germany}
\author{F.~R.~Bouchet\,\orcidlink{0000-0002-8051-2924}}
\affiliation{Sorbonne Universit\'e, CNRS, UMR 7095, Institut d'Astrophysique de Paris, 98 bis bd Arago, 75014 Paris, France}
\author{E.~Camphuis\,\orcidlink{0000-0003-3483-8461}}
\affiliation{Sorbonne Universit\'e, CNRS, UMR 7095, Institut d'Astrophysique de Paris, 98 bis bd Arago, 75014 Paris, France}
\author{M.~G.~Campitiello}
\affiliation{High-Energy Physics Division, Argonne National Laboratory, 9700 South Cass Avenue, Lemont, IL, 60439, USA}
\author{J.~E.~Carlstrom\,\orcidlink{0000-0002-2044-7665}}
\affiliation{Department of Astronomy and Astrophysics, University of Chicago, 5640 South Ellis Avenue, Chicago, IL, 60637, USA}
\affiliation{Kavli Institute for Cosmological Physics, University of Chicago, 5640 South Ellis Avenue, Chicago, IL, 60637, USA}
\affiliation{High-Energy Physics Division, Argonne National Laboratory, 9700 South Cass Avenue, Lemont, IL, 60439, USA}
\affiliation{Enrico Fermi Institute, University of Chicago, 5640 South Ellis Avenue, Chicago, IL, 60637, USA}
\affiliation{Department of Physics, University of Chicago, 5640 South Ellis Avenue, Chicago, IL, 60637, USA}
\author{J.~Carron\,\orcidlink{0000-0002-5751-1392}}
\affiliation{Istituto ricerche solari Aldo e Cele Dacc\`o (IRSOL), Faculty of Informatics, Universit\`a della Svizzera italiana, 6605 Locarno, Switzerland}
\affiliation{Universit\'e de Gen\`eve, D\'epartement de Physique Th\'eorique, 24 Quai Ansermet, CH-1211 Gen\`eve 4, Switzerland}
\author{C.~L.~Chang}
\affiliation{Department of Astronomy and Astrophysics, University of Chicago, 5640 South Ellis Avenue, Chicago, IL, 60637, USA}
\affiliation{Kavli Institute for Cosmological Physics, University of Chicago, 5640 South Ellis Avenue, Chicago, IL, 60637, USA}
\affiliation{High-Energy Physics Division, Argonne National Laboratory, 9700 South Cass Avenue, Lemont, IL, 60439, USA}
\author{P.~M.~Chichura\,\orcidlink{0000-0002-5397-9035}}
\affiliation{Kavli Institute for Cosmological Physics, University of Chicago, 5640 South Ellis Avenue, Chicago, IL, 60637, USA}
\affiliation{Department of Physics, University of Chicago, 5640 South Ellis Avenue, Chicago, IL, 60637, USA}
\author{A.~Chokshi}
\affiliation{Department of Astronomy and Astrophysics, University of Chicago, 5640 South Ellis Avenue, Chicago, IL, 60637, USA}
\author{T.-L.~Chou\,\orcidlink{0000-0002-3091-8790}}
\affiliation{Department of Astronomy and Astrophysics, University of Chicago, 5640 South Ellis Avenue, Chicago, IL, 60637, USA}
\affiliation{Kavli Institute for Cosmological Physics, University of Chicago, 5640 South Ellis Avenue, Chicago, IL, 60637, USA}
\affiliation{National Taiwan University, No. 1, Sec. 4, Roosevelt Road, Taipei 106319, Taiwan}
\author{T.~M.~Crawford\,\orcidlink{0000-0001-9000-5013}}
\affiliation{Department of Astronomy and Astrophysics, University of Chicago, 5640 South Ellis Avenue, Chicago, IL, 60637, USA}
\affiliation{Kavli Institute for Cosmological Physics, University of Chicago, 5640 South Ellis Avenue, Chicago, IL, 60637, USA}
\author{C.~Daley\,\orcidlink{0000-0002-3760-2086}}
\affiliation{Universit\'e Paris-Saclay, Universit\'e Paris Cit\'e, CEA, CNRS, AIM, 91191, Gif-sur-Yvette, France}
\affiliation{Department of Astronomy, University of Illinois Urbana-Champaign, 1002 West Green Street, Urbana, IL, 61801, USA}
\author{T.~de~Haan\,\orcidlink{0000-0001-5105-9473}}
\affiliation{High Energy Accelerator Research Organization (KEK), Tsukuba, Ibaraki 305-0801, Japan}
\author{K.~R.~Dibert}
\affiliation{Department of Astronomy and Astrophysics, University of Chicago, 5640 South Ellis Avenue, Chicago, IL, 60637, USA}
\affiliation{Kavli Institute for Cosmological Physics, University of Chicago, 5640 South Ellis Avenue, Chicago, IL, 60637, USA}
\author{M.~A.~Dobbs}
\affiliation{Department of Physics and McGill Space Institute, McGill University, 3600 Rue University, Montreal, Quebec H3A 2T8, Canada}
\affiliation{Canadian Institute for Advanced Research, CIFAR Program in Gravity and the Extreme Universe, Toronto, ON, M5G 1Z8, Canada}
\author{M.~Doohan}
\affiliation{School of Physics, University of Melbourne, Parkville, VIC 3010, Australia}
\author{D.~Dutcher\,\orcidlink{0000-0002-9962-2058}}
\affiliation{Joseph Henry Laboratories of Physics, Jadwin Hall, Princeton University, Princeton, NJ 08544, USA}
\author{C.~Feng}
\affiliation{Department of Astronomy, University of Science and Technology of China, Hefei 230026, China}
\affiliation{School of Astronomy and Space Science, University of Science and Technology of China, Hefei 230026}
\affiliation{Department of Physics, University of Illinois Urbana-Champaign, 1110 West Green Street, Urbana, IL, 61801, USA}
\author{K.~R.~Ferguson\,\orcidlink{0000-0002-4928-8813}}
\affiliation{Department of Physics and Astronomy, University of California, Los Angeles, CA, 90095, USA}
\affiliation{Department of Physics and Astronomy, Michigan State University, East Lansing, MI 48824, USA}
\author{N.~C.~Ferree\,\orcidlink{0000-0002-7130-7099}}
\affiliation{Kavli Institute for Particle Astrophysics and Cosmology, Stanford University, 452 Lomita Mall, Stanford, CA, 94305, USA}
\affiliation{Department of Physics, Stanford University, 382 Via Pueblo Mall, Stanford, CA, 94305, USA}
\affiliation{California Institute of Technology, 1200 East California Boulevard., Pasadena, CA, 91125, USA}
\author{K.~Fichman}
\affiliation{Kavli Institute for Cosmological Physics, University of Chicago, 5640 South Ellis Avenue, Chicago, IL, 60637, USA}
\affiliation{Department of Physics, University of Chicago, 5640 South Ellis Avenue, Chicago, IL, 60637, USA}
\author{A.~Foster\,\orcidlink{0000-0002-7145-1824}}
\affiliation{Joseph Henry Laboratories of Physics, Jadwin Hall, Princeton University, Princeton, NJ 08544, USA}
\author{S.~Galli}
\affiliation{Sorbonne Universit\'e, CNRS, UMR 7095, Institut d'Astrophysique de Paris, 98 bis bd Arago, 75014 Paris, France}
\author{A.~E.~Gambrel}
\affiliation{Kavli Institute for Cosmological Physics, University of Chicago, 5640 South Ellis Avenue, Chicago, IL, 60637, USA}
\author{A.~K.~Gao}
\affiliation{Department of Physics, University of Illinois Urbana-Champaign, 1110 West Green Street, Urbana, IL, 61801, USA}
\author{F.~Ge\,\orcidlink{0000-0002-3833-8133}}
\affiliation{Kavli Institute for Particle Astrophysics and Cosmology, Stanford University, 452 Lomita Mall, Stanford, CA, 94305, USA}
\affiliation{Department of Physics, Stanford University, 382 Via Pueblo Mall, Stanford, CA, 94305, USA}
\affiliation{California Institute of Technology, 1200 East California Boulevard., Pasadena, CA, 91125, USA}
\affiliation{Department of Physics \& Astronomy, University of California, One Shields Avenue, Davis, CA 95616, USA}
\author{F.~Guidi\,\orcidlink{0000-0001-7593-3962}}
\affiliation{Sorbonne Universit\'e, CNRS, UMR 7095, Institut d'Astrophysique de Paris, 98 bis bd Arago, 75014 Paris, France}
\affiliation{Department of Physics \& Astronomy, University of California, One Shields Avenue, Davis, CA 95616, USA}
\author{S.~Guns}
\affiliation{Department of Physics, University of California, Berkeley, CA, 94720, USA}
\author{N.~W.~Halverson}
\affiliation{CASA, Department of Astrophysical and Planetary Sciences, University of Colorado, Boulder, CO, 80309, USA }
\affiliation{Department of Physics, University of Colorado, Boulder, CO, 80309, USA}
\author{E.~Hivon\,\orcidlink{0000-0003-1880-2733}}
\affiliation{Sorbonne Universit\'e, CNRS, UMR 7095, Institut d'Astrophysique de Paris, 98 bis bd Arago, 75014 Paris, France}
\author{G.~P.~Holder\,\orcidlink{0000-0002-0463-6394}}
\affiliation{Department of Physics, University of Illinois Urbana-Champaign, 1110 West Green Street, Urbana, IL, 61801, USA}
\author{W.~L.~Holzapfel\,\orcidlink{0000-0002-5744-6439}}
\affiliation{Department of Physics, University of California, Berkeley, CA, 94720, USA}
\author{J.~C.~Hood}
\affiliation{Kavli Institute for Cosmological Physics, University of Chicago, 5640 South Ellis Avenue, Chicago, IL, 60637, USA}
\author{A.~Hryciuk}
\affiliation{Kavli Institute for Cosmological Physics, University of Chicago, 5640 South Ellis Avenue, Chicago, IL, 60637, USA}
\affiliation{Department of Physics, University of Chicago, 5640 South Ellis Avenue, Chicago, IL, 60637, USA}
\author{N.~Huang\,\orcidlink{0000-0003-3595-0359}}
\affiliation{Department of Physics, University of California, Berkeley, CA, 94720, USA}
\author{T.~Jhaveri}
\affiliation{Department of Astronomy and Astrophysics, University of Chicago, 5640 South Ellis Avenue, Chicago, IL, 60637, USA}
\affiliation{Kavli Institute for Cosmological Physics, University of Chicago, 5640 South Ellis Avenue, Chicago, IL, 60637, USA}
\author{F.~K\'eruzor\'e}
\affiliation{High-Energy Physics Division, Argonne National Laboratory, 9700 South Cass Avenue, Lemont, IL, 60439, USA}
\author{A.~R.~Khalife\,\orcidlink{0000-0002-8388-4950}}
\affiliation{Sorbonne Universit\'e, CNRS, UMR 7095, Institut d'Astrophysique de Paris, 98 bis bd Arago, 75014 Paris, France}
\author{L.~Knox}
\affiliation{Department of Physics \& Astronomy, University of California, One Shields Avenue, Davis, CA 95616, USA}
\author{K.~Kornoelje}
\affiliation{Department of Astronomy and Astrophysics, University of Chicago, 5640 South Ellis Avenue, Chicago, IL, 60637, USA}
\affiliation{Kavli Institute for Cosmological Physics, University of Chicago, 5640 South Ellis Avenue, Chicago, IL, 60637, USA}
\affiliation{High-Energy Physics Division, Argonne National Laboratory, 9700 South Cass Avenue, Lemont, IL, 60439, USA}
\author{C.-L.~Kuo}
\affiliation{Kavli Institute for Particle Astrophysics and Cosmology, Stanford University, 452 Lomita Mall, Stanford, CA, 94305, USA}
\affiliation{Department of Physics, Stanford University, 382 Via Pueblo Mall, Stanford, CA, 94305, USA}
\affiliation{SLAC National Accelerator Laboratory, 2575 Sand Hill Road, Menlo Park, CA, 94025, USA}
\author{K.~Levy}
\affiliation{School of Physics, University of Melbourne, Parkville, VIC 3010, Australia}
\author{Y.~Li\,\orcidlink{0000-0002-4820-1122}}
\affiliation{Kavli Institute for Cosmological Physics, University of Chicago, 5640 South Ellis Avenue, Chicago, IL, 60637, USA}
\author{A.~E.~Lowitz\,\orcidlink{0000-0002-4747-4276}}
\affiliation{Kavli Institute for Cosmological Physics, University of Chicago, 5640 South Ellis Avenue, Chicago, IL, 60637, USA}
\author{C.~Lu}
\affiliation{Department of Physics, University of Illinois Urbana-Champaign, 1110 West Green Street, Urbana, IL, 61801, USA}
\author{G.~P.~Lynch\,\orcidlink{0009-0004-3143-1708}}
\affiliation{Department of Physics \& Astronomy, University of California, One Shields Avenue, Davis, CA 95616, USA}
\author{T.~J.~Maccarone\,\orcidlink{0000-0003-0976-4755}}
\affiliation{Department of Physics \& Astronomy, Box 41051, Texas Tech University, Lubbock TX 79409-1051, USA}
\author{A.~S.~Maniyar\,\orcidlink{0000-0002-4617-9320}}
\affiliation{Department of Physics, Stanford University, 382 Via Pueblo Mall, Stanford, CA, 94305, USA}
\affiliation{Laboratoire d’Annecy de Physique Theorique (LAPTh), CNRS/USMB, 9 Chemin de Bellevue BP110, Annecy, F-74941, ANNECY CEDEX, France}
\author{E.~S.~Martsen\,\orcidlink{0009-0009-0719-9781}}
\affiliation{Department of Astronomy and Astrophysics, University of Chicago, 5640 South Ellis Avenue, Chicago, IL, 60637, USA}
\affiliation{Kavli Institute for Cosmological Physics, University of Chicago, 5640 South Ellis Avenue, Chicago, IL, 60637, USA}
\author{F.~Menanteau}
\affiliation{Department of Astronomy, University of Illinois Urbana-Champaign, 1002 West Green Street, Urbana, IL, 61801, USA}
\affiliation{Center for AstroPhysical Surveys, National Center for Supercomputing Applications, Urbana, IL, 61801, USA}
\author{M.~Millea\,\orcidlink{0000-0001-7317-0551}}
\affiliation{Department of Physics, University of California, Berkeley, CA, 94720, USA}
\author{J.~Montgomery}
\affiliation{Department of Physics and McGill Space Institute, McGill University, 3600 Rue University, Montreal, Quebec H3A 2T8, Canada}
\author{Y.~Nakato}
\affiliation{Department of Physics, Stanford University, 382 Via Pueblo Mall, Stanford, CA, 94305, USA}
\author{T.~Natoli}
\affiliation{Kavli Institute for Cosmological Physics, University of Chicago, 5640 South Ellis Avenue, Chicago, IL, 60637, USA}
\author{A.~Ouellette\,\orcidlink{0000-0003-0170-5638}}
\affiliation{Department of Physics, University of Illinois Urbana-Champaign, 1110 West Green Street, Urbana, IL, 61801, USA}
\author{Z.~Pan\,\orcidlink{0000-0002-6164-9861}}
\affiliation{Kavli Institute for Cosmological Physics, University of Chicago, 5640 South Ellis Avenue, Chicago, IL, 60637, USA}
\affiliation{High-Energy Physics Division, Argonne National Laboratory, 9700 South Cass Avenue, Lemont, IL, 60439, USA}
\affiliation{Department of Physics, University of Chicago, 5640 South Ellis Avenue, Chicago, IL, 60637, USA}
\author{P.~Paschos}
\affiliation{Enrico Fermi Institute, University of Chicago, 5640 South Ellis Avenue, Chicago, IL, 60637, USA}
\author{K.~A.~Phadke\,\orcidlink{0000-0001-7946-557X}}
\affiliation{Department of Astronomy, University of Illinois Urbana-Champaign, 1002 West Green Street, Urbana, IL, 61801, USA}
\affiliation{Center for AstroPhysical Surveys, National Center for Supercomputing Applications, Urbana, IL, 61801, USA}
\affiliation{NSF-Simons AI Institute for the Sky (SkAI), 172 E. Chestnut St., Chicago, IL 60611, USA}
\author{K.~Prabhu}
\affiliation{Department of Physics \& Astronomy, University of California, One Shields Avenue, Davis, CA 95616, USA}
\author{W.~Quan\,\orcidlink{0009-0002-2589-5501}}
\affiliation{Kavli Institute for Cosmological Physics, University of Chicago, 5640 South Ellis Avenue, Chicago, IL, 60637, USA}
\affiliation{High-Energy Physics Division, Argonne National Laboratory, 9700 South Cass Avenue, Lemont, IL, 60439, USA}
\affiliation{Department of Physics, University of Chicago, 5640 South Ellis Avenue, Chicago, IL, 60637, USA}
\author{A.~Rahlin\,\orcidlink{0000-0003-3953-1776}}
\affiliation{Department of Astronomy and Astrophysics, University of Chicago, 5640 South Ellis Avenue, Chicago, IL, 60637, USA}
\affiliation{Kavli Institute for Cosmological Physics, University of Chicago, 5640 South Ellis Avenue, Chicago, IL, 60637, USA}
\author{C.~L.~Reichardt\,\orcidlink{0000-0003-2226-9169}}
\affiliation{School of Physics, University of Melbourne, Parkville, VIC 3010, Australia}
\author{M.~Rouble}
\affiliation{Department of Physics and McGill Space Institute, McGill University, 3600 Rue University, Montreal, Quebec H3A 2T8, Canada}
\author{J.~E.~Ruhl}
\affiliation{Department of Physics, Case Western Reserve University, Cleveland, OH, 44106, USA}
\author{A.~C.~Silva~Oliveira\,\orcidlink{0000-0001-5755-5865}}
\affiliation{Kavli Institute for Particle Astrophysics and Cosmology, Stanford University, 452 Lomita Mall, Stanford, CA, 94305, USA}
\affiliation{Department of Physics, Stanford University, 382 Via Pueblo Mall, Stanford, CA, 94305, USA}
\affiliation{California Institute of Technology, 1200 East California Boulevard., Pasadena, CA, 91125, USA}
\author{A.~Simpson}
\affiliation{Department of Astronomy and Astrophysics, University of Chicago, 5640 South Ellis Avenue, Chicago, IL, 60637, USA}
\affiliation{Kavli Institute for Cosmological Physics, University of Chicago, 5640 South Ellis Avenue, Chicago, IL, 60637, USA}
\author{J.~A.~Sobrin\,\orcidlink{0000-0001-6155-5315}}
\affiliation{Fermi National Accelerator Laboratory, MS209, P.O. Box 500, Batavia, IL, 60510, USA}
\affiliation{Department of Physics, Villanova University, 800 E Lancaster Ave, Villanova, PA 19085, USA}
\author{A.~A.~Stark}
\affiliation{Center for Astrophysics \textbar{} Harvard \& Smithsonian, 60 Garden Street, Cambridge, MA, 02138, USA}
\author{J.~Stephen}
\affiliation{Enrico Fermi Institute, University of Chicago, 5640 South Ellis Avenue, Chicago, IL, 60637, USA}
\author{C.~Tandoi\,\orcidlink{0000-0002-2077-6004}}
\affiliation{Department of Astronomy, University of Illinois Urbana-Champaign, 1002 West Green Street, Urbana, IL, 61801, USA}
\author{Q.~Taylor\,\orcidlink{0000-0002-6829-4389}}
\affiliation{Department of Physics, Case Western Reserve University, Cleveland, OH, 44106, USA}
\affiliation{CERCA/ISO, Department of Physics, Case Western Reserve University, Cleveland, OH 44106, USA}
\author{C.~Trendafilova\,\orcidlink{0000-0001-5500-4058}}
\affiliation{Center for AstroPhysical Surveys, National Center for Supercomputing Applications, Urbana, IL, 61801, USA}
\affiliation{CERCA/ISO, Department of Physics, Case Western Reserve University, Cleveland, OH 44106, USA}
\author{J.~D.~Vieira\,\orcidlink{0000-0001-7192-3871}}
\affiliation{Department of Astronomy, University of Illinois Urbana-Champaign, 1002 West Green Street, Urbana, IL, 61801, USA}
\affiliation{Department of Physics, University of Illinois Urbana-Champaign, 1110 West Green Street, Urbana, IL, 61801, USA}
\affiliation{Center for AstroPhysical Surveys, National Center for Supercomputing Applications, Urbana, IL, 61801, USA}
\author{A.~G.~Vieregg\,\orcidlink{0000-0002-4528-9886}}
\affiliation{Department of Astronomy and Astrophysics, University of Chicago, 5640 South Ellis Avenue, Chicago, IL, 60637, USA}
\affiliation{Kavli Institute for Cosmological Physics, University of Chicago, 5640 South Ellis Avenue, Chicago, IL, 60637, USA}
\affiliation{Enrico Fermi Institute, University of Chicago, 5640 South Ellis Avenue, Chicago, IL, 60637, USA}
\affiliation{Department of Physics, University of Chicago, 5640 South Ellis Avenue, Chicago, IL, 60637, USA}
\author{A.~Vitrier\,\orcidlink{0009-0009-3168-092X}}
\affiliation{Sorbonne Universit\'e, CNRS, UMR 7095, Institut d'Astrophysique de Paris, 98 bis bd Arago, 75014 Paris, France}
\author{Y.~Wan\,\orcidlink{0009-0007-0072-0445}}
\affiliation{Department of Astronomy, University of Illinois Urbana-Champaign, 1002 West Green Street, Urbana, IL, 61801, USA}
\affiliation{Center for AstroPhysical Surveys, National Center for Supercomputing Applications, Urbana, IL, 61801, USA}
\author{N.~Whitehorn\,\orcidlink{0000-0002-3157-0407}}
\affiliation{Department of Physics and Astronomy, Michigan State University, East Lansing, MI 48824, USA}
\author{W.~L.~K.~Wu\,\orcidlink{0000-0001-5411-6920}}
\affiliation{Kavli Institute for Particle Astrophysics and Cosmology, Stanford University, 452 Lomita Mall, Stanford, CA, 94305, USA}
\affiliation{SLAC National Accelerator Laboratory, 2575 Sand Hill Road, Menlo Park, CA, 94025, USA}
\affiliation{California Institute of Technology, 1200 East California Boulevard., Pasadena, CA, 91125, USA}
\author{M.~R.~Young}
\affiliation{Kavli Institute for Cosmological Physics, University of Chicago, 5640 South Ellis Avenue, Chicago, IL, 60637, USA}
\affiliation{Fermi National Accelerator Laboratory, MS209, P.O. Box 500, Batavia, IL, 60510, USA}
\author{J.~A.~Zebrowski\,\orcidlink{0000-0003-2375-0229}}
\affiliation{Department of Astronomy and Astrophysics, University of Chicago, 5640 South Ellis Avenue, Chicago, IL, 60637, USA}
\affiliation{Kavli Institute for Cosmological Physics, University of Chicago, 5640 South Ellis Avenue, Chicago, IL, 60637, USA}
\affiliation{Fermi National Accelerator Laboratory, MS209, P.O. Box 500, Batavia, IL, 60510, USA}
\collaboration{SPT-3G Collaboration}
\noaffiliation

\date{\today}

\begin{abstract}
We search for anisotropic cosmic birefringence using two years of data from the SPT-3G instrument on the South Pole Telescope.
Anisotropic cosmic birefringence refers to a theorized dependence of the refractive index of the vacuum on the polarization of light that varies across the Universe in a statistically isotropic manner. 
We present a measurement of the power spectrum of the rotation angle $\alpha$ in six bins across the angular multipole range $20 < L < 2048$, finding no evidence for cosmic birefringence with the spectrum consistent with zero. 
At 95\% confidence, we set a new best upper limit on the amplitude of a scale-invariant birefringence power spectrum of \upperscaleinvar. 
Additionally, we place upper limits on a scale-dependent white noise model.
We translate the scale-invariant upper limit to constraints on the axion coupling constant and primordial magnetic field strength, finding 95\% confidence limits of \upperscaleginvar{} and \upperscaleBinvar{} respectively. We expect these limits to improve several-fold with the complete SPT-3G survey.
\end{abstract}

\maketitle

\section{\label{sec:level1}Introduction}

Modern cosmic microwave background (CMB) experiments, such as SPT-3G \citep{sobrin22}, Simons Observatory \citep{galitzki18} and BICEP Array \citep{hui18}, are producing ever-more precise measurements of the polarization anisotropies of the CMB in the quest to better understand conditions in the very early Universe. 
These polarization measurements also facilitate tests for more exotic physics, such as cosmic birefringence \citep{komatsu2022}, which would rotate the polarization plane of photons as they travel through the cosmos. Such a birefringence signal could indicate the presence of axion-like particles (ALPs) or primordial magnetic fields (PMFs), which we expand upon in \S\ref{sec:theory}. Axion-like particles could create an achromatic isotropic \citep{finelli09, carroll98}, anisotropic \citep{lue99, caldwell11}, or time-varying birefringence signal \citep{fedderke19}, while primordial magnetic fields could create a chromatic, static or nearly static isotropic or anisotropic birefringence signal \citep{kosowsky96, harari97}.

Searches have been conducted for static birefringence, both global (isotropic) and anisotropic, and for time-varying birefringence. 
There have been suggestions of a non-zero global rotation signal in \planck{} satellite data \citep{Minami_2020, Eskilt_2022a, eskilt_2022b}, though this result depends on the use of the Milky Way as a polarization angle calibrator. 
An analysis of ACT DR6 data employed instrument optics–model priors to separate polarization angle systematics from a potential isotropic birefringence signal and reported a non-zero global rotation angle, although residual systematics cannot yet be ruled out as the cause of this rotation \citep{diegopalazuelos2025}. 
There has also been a low-significance suggestion of a non-zero time-varying signal in observation of Tau A with \polarbear{} \citep{adachi24}, but the detection level conflicts with the other reported upper limits on the time-varying signal from CMB polarization with the South Pole Telescope \citep{ferguson22}, BICEP/\textit{Keck} \citep{Ade_2021, Ade_2022}, and POLARBEAR \citep{Adachi2023} and was not detected in a follow-up search with the Simons Array experiment \citep{adkins2025}. 
Previous measurements of static, anisotropic birefringence have been consistent with zero \citep{bianchini20b, namikawa20, polarbear15, bicepkeck_XVII, Zagatti_2024, namikawa25, lonappan2025}.

In this work, we improve upon a previous measurement of the anisotropic birefringence rotation angle power spectrum from the SPTpol instrument on the South Pole Telescope (SPT) \citep{bianchini20b} using data from the latest SPT camera, SPT-3G. 
We use two years of 150\,GHz data from observations of the Main 1500\,\sqdeg{} field during the 2019 and 2020 observing seasons. 
We use the measured rotation angle power spectrum to place improved limits on both a scale-invariant and a white noise rotation power spectrum.

The outline of this work is as follows. 
We discuss the the theory of cosmic birefringence in \S\ref{sec:theory}. 
We lay out the data and simulations used by the analysis framework in \S\ref{sec:data}, before describing the analysis methods in \S\ref{sec:analysis}. We discuss analysis validation in \S\ref{sec:validation}.
We present the bandpowers and parameter fitting in \S\ref{sec:results}, before concluding in \S\ref{sec:conclusion}.

\section{\label{sec:theory}Theoretical Framework of Cosmic Birefringence}

Cosmic birefringence refers to a rotation of the linear polarization plane of photons as they travel through the universe. 
This work focuses on the anisotropic cosmic birefringence case, where the rotation angle $\alpha(\hat{n})$ varies with sky position $\hat{n}$ but does not temporally vary (or varies on longer timescales than we have measured CMB polarization). 
As noted in the introduction, there have also been searches for global birefringence (where the rotation is not position-dependent) and time-varying birefringence.

Cosmic birefringence is  predicted by extensions of the standard cosmological model that include pseudoscalar fields such as axion-like particles (ALPs) \cite{Carroll1990,Marsh2016}. 
These theories have emerged in an effort to explain the dark sector or inflation. 
The interaction of an ALP field $a$ with the electromagnetic field $\mathcal{F}_{\mu \nu}$ modifies the Lagrangian with a Chern-Simons term \cite{carroll90}:
\begin{equation}
    \mathcal{L}_{CS} \supset \frac{g_{\alpha\gamma}}{4}a\mathcal{F}_{\mu \nu} \widetilde{\mathcal{F}}^{\mu \nu}
\end{equation}
where $g_{\alpha\gamma}$ is the axion coupling constant with mass dimension $-1$ and $a$ is the nearly massless axion-like pseudoscalar field. 

The parity-violating nature of this term introduces different refractive indices for right- and left-handed circular polarization states of light. 
As linear polarization is the superposition of right and left-handed circular states with an angle proportional to the phase difference between the two, a difference in the refractive indices causes a change in the phase difference and hence the linear polarization angle. 
Scalar perturbations in the early universe are expected to produce pure $E$-mode polarization patterns. 
The rotation from cosmic birefringence would convert the initially pure $E$-mode pattern into a mixture of $E$- and $B$-modes. 
This $E$-to-$B$ conversion correlates the $E$ and $B$ polarization fields, and this correlation can be used to reconstruct the rotation angle field across the sky. 
The change in polarization angle is proportional to the integrated change in the field over the light trajectory ($\Delta a$) \citep{Harari1992}:
\begin{equation}
    \alpha (\hat{\mathbf{n}})= \frac{g_{\alpha\gamma}}{2} \Delta a.
\end{equation}

\subsection{\label{subsec:scaleinvar}Scale-Invariant Anisotropic Birefringence}

If the pseudoscalar field $a$ is position-dependent, the resulting rotation angles will be anisotropic. 
Such fields are often introduced as a massless scalar field during inflation, where they are expected to produce a scale-invariant rotation spectrum at large scales \citep{caldwell11}:
\begin{equation}\label{eqn:axion}
    \frac{L(L+1)}{2\pi}C^{\alpha\alpha}_L = \left( \frac{H_I g_{\alpha\gamma}}{4\pi} \right)^2.
\end{equation}
Thus the scale-invariant level of the rotation angle power spectrum is expected to be proportional to the square of the product of the axion coupling constant $g_{\alpha\gamma}$ and Hubble parameter during the inflationary era $H_I$. 

If they exist, primordial magnetic fields (PMFs) would have a similar effect on CMB polarization due to the resulting Faraday rotation in the plasma of the early Universe \citep{Kosowsky1996,Harari_1997},
\begin{equation}
    \alpha (\hat{\mathbf{n}})=\frac{3}{16 \pi^2 e \,\nu ^2} \int d\boldsymbol{l} \cdot \dot{\tau} \boldsymbol{B},
\end{equation}
where $\nu$ is the observed frequency, $\dot{\tau}$ is the differential optical depth, and $\boldsymbol{B}$ is the co-moving magnetic field strength. The expected scale-independent spectrum of primordial magnetic field strength yields a scale-invariant rotation angle power spectrum:
\begin{equation}\label{eqn:pmf}
\frac{L(L+1)}{2\pi} C_{L}^{\alpha\alpha} = 3.6\times 10^{-8} \left( \frac{\nu}{\rm 150\,GHz}\right)^{-4} \left( \frac{\bpc}{\rm 1\,nG}\right)^2 .
\end{equation} 
Here, \bpc{} is the RMS magnetic field strength on 1\,Mpc scales.
Unlike a Chern-Simons coupling, Faraday rotation is frequency-dependent, with the rotation angle proportional to the inverse square of the photon frequency.  
If birefringence is detected, this frequency dependence could be used to distinguish between the two scenarios. 

As most models of anisotropic cosmic birefringence predict a scale-invariant spectrum, we define our primary reference power spectrum as 
\begin{equation}\label{eqn:acb}
    \frac{L(L+1)}{2\pi} C_{L}^{\alpha\alpha} = A_{\textrm{CB}} \times 10^{-4}\ [\text{rad}^2],
\end{equation}
where $A_{\textrm{CB}}$ can be related to the physical parameters described above. We normalize by $10^{-4}$ for consistency with previous publications.

\subsection{\label{subsec:scaledep}Scale-Dependent Anisotropic Birefringence}

If the aforementioned axion-like field evolves after recombination, it would produce a scale-dependent anisotropic birefringence signal \citep{greco2023, ferreira2024}.                  
A well-explored theory for such an evolution is the collapse of an axion string or string-wall network \citep{jain2022}. 
Peccei-Quinn spontaneous symmetry breaking in the early universe leads to the formation of a topological defect network consisting of one-dimensional axion strings of high-energy density. \citep{Kibble_1976}. At late times, axion domain walls can form connecting the axion strings, and collectively these structures are called a string-wall network.
For ultralight axion masses this string-wall network can survive past recombination, delaying its decay into axion particles. 

Because of the finite correlation length of the axion field, limited by causality considerations, at least one string is expected to be present per Hubble volume. Each Hubble volume may have arbitrarily many strings, but simulations show the most likely range to be one to ten strings per Hubble volume \citep{kawasaki2015, Gorghetto_2018, Lopez-Eiguren_2017, Yin_2023, buschmann2022, benabou2025}. 
These strings continuously radiate energy in the form of axions, but considerable birefringence is expected to be created from the axion string network itself, with a photon passing through a string geometry (e.g. a string loop) accumulating more polarization rotation than a photon interacting with an axion particle \citep{jain2021, Agrawal_2020}.

The string-wall network grows with the Hubble scale and will continue to produce birefringence until the network collapses. 
Collapse occurs when the cosmic temperature drops to the axion mass scale ($m_a \sim 3H$ where $H$ is the Hubble parameter); the domain wall tension pulls the strings apart, causing a rapid destruction of the topological defects.
In this scenario the birefringence power spectrum follows a white noise spectrum below the collapse multipole $L_c$, which is set by the number of strings per Hubble volume at recombination and the string network size at collapse. 
The white noise spectrum arises from decorrelation of the signal on large angular scales ($L < L_c$) as the universe expands.
In the case of network collapse just after recombination and one string per Hubble volume, one expects a collapse multipole of $L_c \sim 200$. 
In the case of network collapse just after recombination and ten strings per Hubble volume, one expects a collapse multipole of one tenth the Hubble volume today, or $L_c\sim2000$. 
If network collapse occurs at a later redshift when strings have grown, the collapse multipole will be correspondingly decreased.

We model this collapsing string-wall network as the scale-dependent white noise spectrum
\begin{equation}\label{eqn:whitenoise}
    \frac{L(L+1)}{2\pi}C^{\alpha\alpha}_L = A_{\textrm{wn}}L^2 \times 10^{-4}\ [\text{rad}^2].
\end{equation}
We constrain the white noise amplitude $A_\textrm{wn}$ over our full multipole range of $20 < L < 2048$. We do not separately constrain this model for low $L_c$, as the scale-invariant model discussed above serves as a robust test of birefringence power on large scales. 
This model is a simplified version of the collapsing string-wall model introduced in \citet{jain2022} which probes axion masses of $\sim 10^{-28} - 10^{-32}$ eV. The collapsing string-wall network is one of many possible axion network configurations, and demonstrates the benefit of high-$L$ constraining power in non-scale-invariant models.

\section{Data}\label{sec:data}

We use the SPT-3G D1 150 GHz dataset, which consists of the 2019 and 2020 observing seasons with the SPT-3G instrument on the South Pole Telescope over the 1500\,\sqdeg{} SPT-3G Main field. 
SPT-3G has been used to survey over 10,000 deg$^2$, including additional Summer and Wide fields, however the Main field has significantly lower polarization noise levels than the rest of the survey \citep{prabhu24}, and thus has the highest sensitivity to searches for cosmic birefringence.
The SPT-3G survey is expected to finish in 2028,  with nine years of observations on the SPT-3G Main field, thus we expect significant improvements in sensitivity for the full dataset. We limit this analysis to the 150 GHz data, the lowest-noise SPT-3G dataset, and leave the inclusion of 95 and 220 GHz to the future full-depth analysis.

A detailed description of the processing of the time-ordered data and map-making can be found in \citet{camphuis25} and  \citet{quan26}. 
Briefly, we band-pass filter the time-ordered data to reduce both high- and low-frequency noise before binning the data into maps.  
We use \texttt{HEALpix} \citep{gorski05} Nside=8192 maps for this analysis. The individual full-survey maps are coadded with inverse-variance weighting to form the data map used in the analysis. 
We also generate `signflip' noise maps by splitting the map set into random halves and multiplying one half by negative one. We then coadd the full set to null the signal. 
These signflip maps provide a realistic estimate of the instrumental noise in the map.

\subsection{Simulations}\label{subsec:sim}

We simulate the effects of the filtering applied to the time-ordered data and map-making process using end-to-end simulations. 
For these simulations, we resample noiseless sky maps based on where each detector is pointing at each time sample, and run these simulated time-ordered data through the same filtering and map-making steps as the real data. 
As these are noiseless, we simulate a random 5\% of the full data volume to reduce the computational load. 
We have checked that the recovered birefringence power spectrum has converged at this data fraction. 
When needed, we add noise to the simulations using the signflip noise maps described above. 
The lensed CMB anisotropies are drawn from a best-fit \planck{} 2018 \lcdm{} cosmology. 

We use several sets of simulations in this analysis. 
\begin{enumerate}
\item 200 realizations of the unrotated lensed CMB sky, with foregrounds and instrumental noise. 
\item 200 realizations of the rotated lensed CMB sky, with foregrounds and instrumental noise. The lensed primary CMB is the same as in set 1.
\item 50 realizations of the unrotated lensed CMB sky, with foregrounds and instrumental noise. The lensed primary CMB differs from sets 1 and 2.
\item 24 pairs of simulations of the rotated lensed CMB sky, without foregrounds or instrumental noise. Each pair has a different lensed CMB sky but the same rotation field. 
\end{enumerate}
The rotated simulations are made from the unrotated simulations with realizations of the rotation angle maps for a scale-invariant $A_{\textrm{CB}} = 0.1$ rotational angle power spectrum. Simulation set (2) is used only to validate our rotation angle reconstruction.

\subsection{Initial processing}

The measured Q/U maps are masked to remove bright point sources (150\,GHz temperature flux above 250 mJy). The area loss from masking these 33 point sources is $<0.6\%$ of the map.
An edge taper of 30 arcmin is applied around both the map edges and these bright sources. 
The masked maps are spherical-harmonic transformed and used to derive harmonic-space estimates of the $E$ and $B$-mode fields, $E_{\ell m}$ and $B_{\ell m}$.
These are corrected by the best-fit polarization calibration and polarized beam from \citet{camphuis25}, which used the same maps. 
A similar correction is applied to the signflip maps and simulated maps. 
However we correct the simulated maps by the beam used when creating the simulations. 
These source-masked, calibrated and beam-corrected maps are the inputs to the birefringence pipeline.

Given that the map-making process is lossy, we do not expect to recover a perfect estimate of all sky modes. 
We calculate a 2D transfer function for $E_{\ell m}$ and $B_{\ell m}$ and zero all $\ell, m$ regions that are heavily impacted by the high-pass filter applied to the time-ordered data in map-making and our scanning pattern. In detail, this means we do not use modes $m < 120$ and $\ell < 300$, which have transfer function values of $\lessapprox 0.4$. 
Due to the field location, the maps do not measure modes with $\ell \sim m$ \citep[see][]{aylor17} and we also exclude these modes from the analysis. 
Rather than applying a transfer function correction to $E$ and $B$ in $a_{\ell m}$ space and applying another one at the power spectrum stage, we correct for all loss after the $C_L^{\alpha \alpha}$ power spectrum calculation. 
This correction is described in \S\ref{sec:analysis}.

\section{\label{sec:methods} Methods}

\subsection{\label{sec:analysis} Cosmic Birefringence Angle Reconstruction and Power Spectrum Estimation}

We now describe the steps necessary to reconstruct the rotation angle anisotropy from spherical harmonic transforms of the data and simulated maps presented in \S\ref{sec:data}, and to estimate the power spectrum of the resulting rotation angle maps. 
The analysis closely follows the approach used with SPTpol data by \citet{bianchini20b}, with the exception of using spherical harmonic transforms instead of Fourier transforms. This spherical harmonic ``curved-sky" method is adapted from \citet{omori2026} and \citet{namikawa20}, and avoids flat-sky projection effects over the larger SPT-3G survey area.  

Similar to the deflection angle effects from CMB lensing, the polarization angle rotation due to cosmic birefringence correlates previously independent modes across the sky. 
The induced off-diagonal mode-mode correlation structure can be used to reconstruct the rotation angle field $\alpha(\hat{n})$ using a quadratic estimator \citep{kamionkowski09,yadav09,gluscevic09,namikawa17}.
The quadratic estimator combines pairs of the $E$- and $B$-mode maps in spherical harmonic space. We follow the formalism and implementation of \citet{namikawa20} and write the unnormalized quadratic estimator of $\alpha$ as

\be
\bar{\alpha}^{*}_{\bf{L}}
=
\sum_{\ell \ell' m m'}
\begin{pmatrix}
\ell & \ell' & L \\
m & m' & M
\end{pmatrix}
f^{\alpha}_{\ell L \ell'}
\, \bar{E}_{\Bell} \, \bar{B}_{\Bell'}
\label{eq:alpha_est_bias}
\ee 

where the weight function is defined as 
\be
f^{\alpha}_{\ell L \ell'}
=
- W^{-}_{\ell' L \ell} \, \tilde{C}^{EE}_{\ell}
\ee
and where
\begin{equation}
\begin{aligned}
        W^{\pm}_{\ell_1 \ell_2 \ell_3}
        =
        \pm 2 \, \zeta^{\mp}_{\ell_1 \ell_2 \ell_3}
        \, p^{\mp}_{\ell_1 \ell_2 \ell_3}
        \sqrt{
        \frac{
        (2\ell_1+1)(2\ell_2+1)(2\ell_3+1)
        }{4\pi}
        }
        \\
        \begin{pmatrix}
        \ell_1 & \ell_2 & \ell_3 \\
        -2 & 0 & 2
        \end{pmatrix}.
\end{aligned}
\end{equation}
Here, $\tilde{C}^{EE}_{\ell}$ is the lensed E-mode power spectrum. $\bar{E}_{ \Bell}$ and $ \bar{B}_{\Bell'}$ denote the observed multipoles diagonally filtered by their inverse variance, $\zeta^+ = 1$, $\zeta^- = \mathrm{i}$, and $p^\pm_{\ell_1 \ell_2 \ell_3} = [1 \pm (-1)^{\ell_1 + \ell_2 + \ell_3}]/2$ is a parity indicator.
To simplify the indices, we have used $\Bell$ to represent ($\ell, m$) of the spherical harmonic transform of a map, and ${\bf L}$ to represent ($L, M$) of the spherical harmonic transform of the angle field. 
The estimation of the $E$ and $B$ fields is described in the last section. 

As with the quadratic estimator for gravitational lensing, the reconstruction from Eqn.~\ref{eq:alpha_est_bias} yields a biased estimate of the true $\alpha$ field. 
We must correct it for several additive and multiplicative biases in order to make an unbiased estimate of the cosmic rotation power spectrum. 

First we subtract a mean-field correction, $\left\langle \bar{\alpha}_{\bf L} \right\rangle$,  from the $\alpha$ maps in harmonic space. 
This mean field is due to statistically anisotropic effects in the map, such as noise variations or mode-coupling induced by the apodization mask. 
The mean-field correction is calculated by taking the average of the reconstructed $\alpha$ maps, $\bar{\alpha}_{\bf L}$,  for 50 of the lensed, unrotated simulations from simulation set (1) in \S\ref{subsec:sim}: 
\be
\left\langle \bar{\alpha}_{\bf L} \right\rangle = \frac{1}{N}\sum \bar{\alpha}_{\bf L}.
\ee
Second, we apply a multiplicative correction, $\mathcal{R}_{L}$ to the  $\alpha$ maps in harmonic space to get an unbiased estimate of the rotation map, $\hat{\alpha}_{\bf L}$ :

\be
\label{eq:alpha_est}
\hat{\alpha}_{\bf L}
=
\mathcal{R}_L
\left(
\bar{\alpha}_{\bf L}
-
\left\langle \bar{\alpha}_{\bf L} \right\rangle
\right)
\ee

This correction is calculated in two parts. 
First we make an analytic estimate of the effects of filtering, using

\be
(\mathcal{R}^{\textrm{an}}_L)^{-1}
=
\frac{1}{2L+1}
\sum_{\ell \ell'}
\frac{
\left( f^{\alpha}_{\ell L \ell'} \right)^2
}{
\hat{C}^{EE}_{\ell}
\hat{C}^{BB}_{\ell'}
}.
\ee
Here, $\hat{C}^{XX}_{\ell}$ are the sum of the power spectrum of the simulated lensed $E$ or $B$ CMB anisotropies described in \S\ref{subsec:sim} and the power spectrum of the signflip noise maps described in \S\ref{sec:data}.
Second, we calculate a multiplicative correction $\mathcal{R}_{L}^{\textrm{MC}}$ to the analytic estimate by calculating the correlation of the input to output rotation fields in the rotated sims of simulation set (2) in \S\ref{subsec:sim}. 
Unlike \cite{bianchini20b}, we apply the total correction $\mathcal{R}_{L} = \mathcal{R}_{L}^{\textrm{an}} \mathcal{R}_{L}^{\textrm{MC}}$. 

The power spectrum $C_{L}^{\hat{\alpha}\hat{\alpha}}$ of the rotation map will be biased high by three additive bias terms. 
We must subtract these terms from the estimate to recover an unbiased estimate of the rotation power spectrum, $\hat{C}^{\alpha\alpha}_L$:
\be
\hat{C}^{\alpha\alpha}_L = C_L^{\hat{\alpha}\hat{\alpha}} - N_L^{(0)}  - N_L^{\rm lens} - N_L^{(1)}.
\ee
The three terms are the disconnected bias, $N_L^{(0)}$, the lensing bias, $N_L^{\rm lens}$, and the connected bias $N_L^{(1)}$.
We discuss each term and how it is estimated in the following paragraphs. It also possible for this measurement to be biased by post-Born lensing curl $B$-modes \citep{robertson2025}, but this is negligible at our current sensitivity.

The largest bias is the disconnected bias term, or $N^{(0)}_L$. 
We use two estimates for the disconnected bias, depending on whether the real data is involved. 
For simulations only, the disconnected bias is estimated as
\be\label{eqn:n0}
N^{(0)}_L = \langle \hat{C}_L^{ijij} + \hat{C}_L^{ijji} \rangle,
\ee
where $\hat{C}_L^{ijij}$ and $\hat{C}_L^{ijji}$ represent cross-spectra between two unrotated (simulation set (1)) $\alpha$ maps constructed from simulation pairs $(i, j)$ with $i = j+1$. 
For the data, we calculate the realization-dependent (RD disconnected bias $N^{(0),\rm RD}_L$ using the algorithm introduced by \citet{namikawa17} to be more robust to any potential mismatch between the data and simulations:
\be
N^{(0),\rm RD}_L = \langle \hat{C}_L^{didi} + \hat{C}_L^{diid} + \hat{C}_L^{idid} + \hat{C}_L^{iddi} \rangle - N^{(0)}_L.
\ee
Here $\hat{C}_L^{didi}$ is the cross-spectrum between the $\alpha$ maps from the data and the $i^{\rm th}$ simulation of simulation set (1), while $N_{L}^{(0)}$ is as defined for Eqn.~\ref{eqn:n0}. 

The second largest bias is the contribution from the lensing-induced trispectrum  \citep{namikawa17}. 
We estimate the lensing bias using lensed, unrotated sims from simulation sets (1) and (3) according to
\bea
N^{\rm lens}_L &=& \langle \hat{C}_L^{iiii}\rangle - N^{(0)}_L.
\eea
The autospectrum of the rotation map of one lensed, unrotated simulation is $\hat{C}_L^{iiii}$, and we calculate this term using simulation set (3).

The connected bias term, $N_L^{(1)}$, only applies to rotated sims. 
It will be zero in the absence of cosmic birefringence.
To be conservative,  we do not subtract $N_L^{(1)}$ from the data when estimating constraints in later sections. 
However, we do subtract the connected bias for pipeline tests to confirm that the pipeline recovers an unbiased estimate of the input power spectrum used in test simulations $C_{L}^{\alpha\alpha, sim}$.
 Following \citet{story15}, we estimate the connected bias as
 \be
N^{(1)}_L = \langle \hat{C}_L^{i'ii'i} + \hat{C}_L^{ii'i'i} - \hat{C}_L^{ijij} - \hat{C}_L^{ijji} \rangle,
\ee
where the $ii'$ pair of simulations share the same rotation angle field $\alpha$ but have different realizations of the lensed CMB polarization anistropies (simulation set (4)). 
As the connected bias term is small and only used in pipeline tests,  we use a smaller set of 24 such simulation pairs to estimate the connected bias in order to reduce the computational cost.

\subsection{Likelihood}

We follow the Hamimeche-Lewis formalism \citep{hamimeche08} to compare the measured rotation angle power spectrum to theoretical models. 
The likelihood code and associated data files are available on the SPT \footnote{\url{https://pole.uchicago.edu}} and LAMBDA \footnote{\url{https://lambda.gsfc.nasa.gov/}} websites. 
We derive posteriors on the model parameters using  Markov chain Monte Carlos (MCMCs), as implemented in the Cobaya package \citep{torrado21}. 

\subsection{\label{sec:validation}Analysis Validation}

We validate the analysis pipeline by inputting sims with a known birefringence power spectrum and verifying the expected output is returned. We perform this test for both the rotated (birefringence signal) and unrotated (no birefringence signal) cases. For the rotated test we include the connected bias term (N1), which is neglected in our analysis of real data.

We vary several parameters of our analysis in order to validate our results and apply these changes to both the sims and real data. First, we change our source masking in two ways: we decrease our source masking threshold from 250 mJy to 100 mJy, and increase our masked source size from 30 arcmin to 60 arcmin. Both of these changes result in a $<0.2\sigma$ shift from our baseline bandpowers, validating that we are not generating significant mode-mixing through masking. We then test changes to our cuts in both $\ell_{\textrm{max}}$ and $m_{\textrm{min}}$ (the maximum or minimum values used in ${l,m}$ space) to test for effects from unmasked point sources and filtering, respectively. With a baseline of $\ell_{\textrm{max}} = 3300$ and $m_{\textrm{min}} = 120$, we repeated our analysis with three ${\ell,\ m}$ configurations: $\ell_{\textrm{max}}=3000,\ m_{\textrm{min}}=120$; $\ell_{\textrm{max}}=3600,\ m_{\textrm{min}}=120$; and $\ell_{\textrm{max}}=3300,\ m_{\textrm{min}} = 200$. We find all bandpowers shift $0.4\sigma$ or less from baseline for the above configurations and therefore assume our results to be robust to reasonable changes in $\ell\ \text{and}\ m$.

\section{\label{sec:results}Results}

\subsection{\label{sec:bandpowers}Cosmic Birefringence Bandpowers}

\begin{figure}
\begin{center}
    \includegraphics[width=\linewidth]{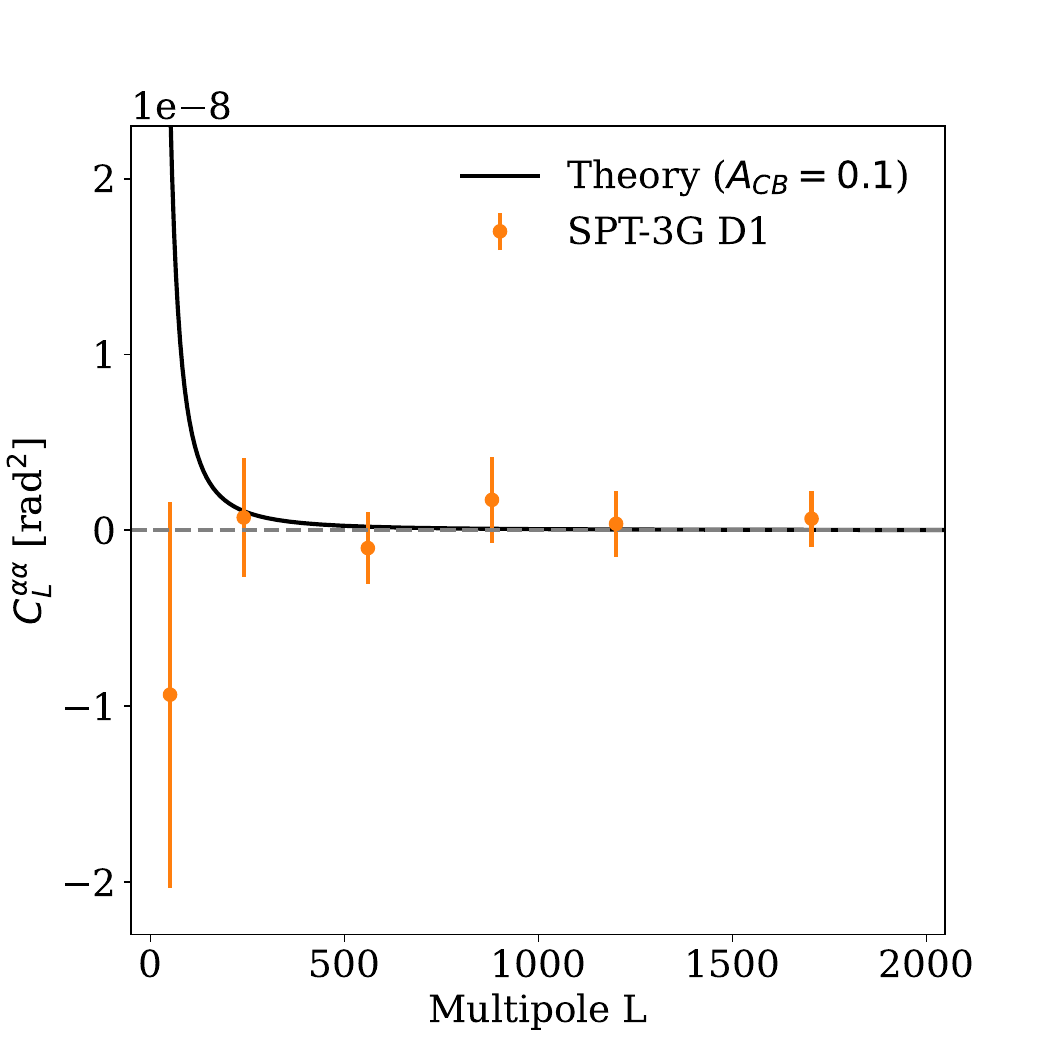}
    \caption{The optimally weighted bandpowers for the rotation angle power spectrum measured with the SPT-3G D1 dataset.  
    The data are consistent with zero. 
    The black line shows the theoretical prediction   for a scale-invariant spectrum with $A_{\textrm{CB}} = 0.1$.
    }
    \label{fig:spectrum}
\end{center}
\end{figure}

We measure a rotation angle power spectrum consistent with zero over the multipole range $20 < L < 2048$, as shown in Fig.~\ref{fig:spectrum}.
We report two sets of bandpowers: one using a uniform weighting of all modes within the bin, and one using the optimal weighting as a function of $\ell$ for a scale-invariant power spectrum. These are tabulated in Table \ref{tab:bandpowers}.
The optimally weighted bandpowers lead to significantly tighter limits on scale-invariant power, as discussed in \S\ref{subsec:upperlim}.
We plot the weighted set of bandpowers in Fig.~\ref{fig:spectrum}.
The data provide no support for a non-zero birefringence power spectrum. 

\begin{figure*}
\begin{center}
    \includegraphics[width=\linewidth]{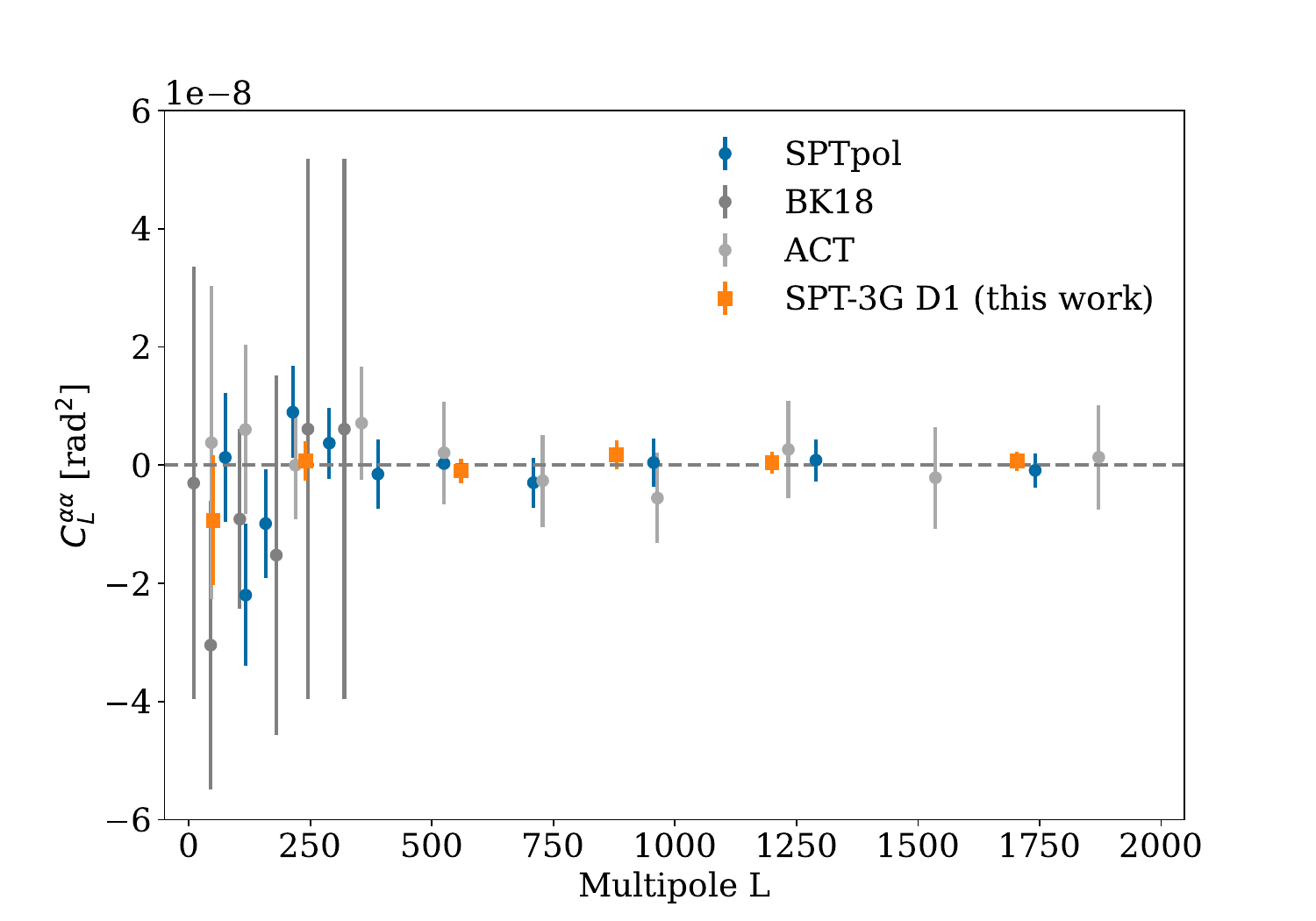}
    \caption{Comparisons of bandpowers from BK18 \citep{ade23}, ACT \citep{namikawa20}, SPTpol \citep{bianchini20b}, and SPT-3G D1 (this work, weighted bandpowers). Zero is marked by the dashed line. SPT-3G D1 significantly improves constraints on the rotation power spectrum $C_{L}^{\alpha\alpha}$.
    }
    \label{fig:spectrumcomp}
\end{center}
\end{figure*}

We compare the SPT-3G measurement to other published anisotropic birefringence measurements from SPTpol \citep{bianchini20b}, BICEP/\textit{Keck} \citep{ade23}, and ACT  \citep{namikawa20} in Fig.~\ref{fig:spectrumcomp}. 

\begin{table}[htbp]
\centering
\caption{Cosmic birefrigence bandpowers from SPT-3G. $C_{L,\textrm{unw}}^{\alpha\alpha}$ denotes the unweighted bandpowers, and $C_{L,\textrm{w}}^{\alpha\alpha}$ denotes the bandpowers with optimal scale-invariant weighting applied.}

\label{tab:bandpowers}
\begin{tabular}{ccccc}

[$L_{min}$ & $L_{max}$] & $L$ & $C_{L,\textrm{unw}}^{\alpha\alpha}$ [$10^{-9}$ rad$^2$]  & $C_{L,\textrm{w}}^{\alpha\alpha}$ [$10^{-9}$ rad$^2$] \\ \hline
20 & 79 & 50 & $-6.16\pm7.62$  & $-9.35\pm11.00$ \\
80 & 399 & 240 & $4.51\pm1.71$  & $0.72\pm3.36$  \\
400 & 719 & 560 & $-1.46\pm2.05$  & $-1.01\pm2.03$  \\
720 & 1039 & 880 & $1.38\pm2.40$  & $1.73\pm2.45$  \\
1040 & 1359 & 1200 & $0.45\pm1.87$  & $0.37\pm1.88$  \\
1360 & 2048 & 1700 & $0.42\pm1.60$  & $0.66\pm1.59$  \\ \hline
\end{tabular}
\end{table}

\subsection{Upper Limits on Cosmic Birefringence}\label{subsec:upperlim}

\begin{figure}
\begin{center}
    \includegraphics[width=\linewidth]{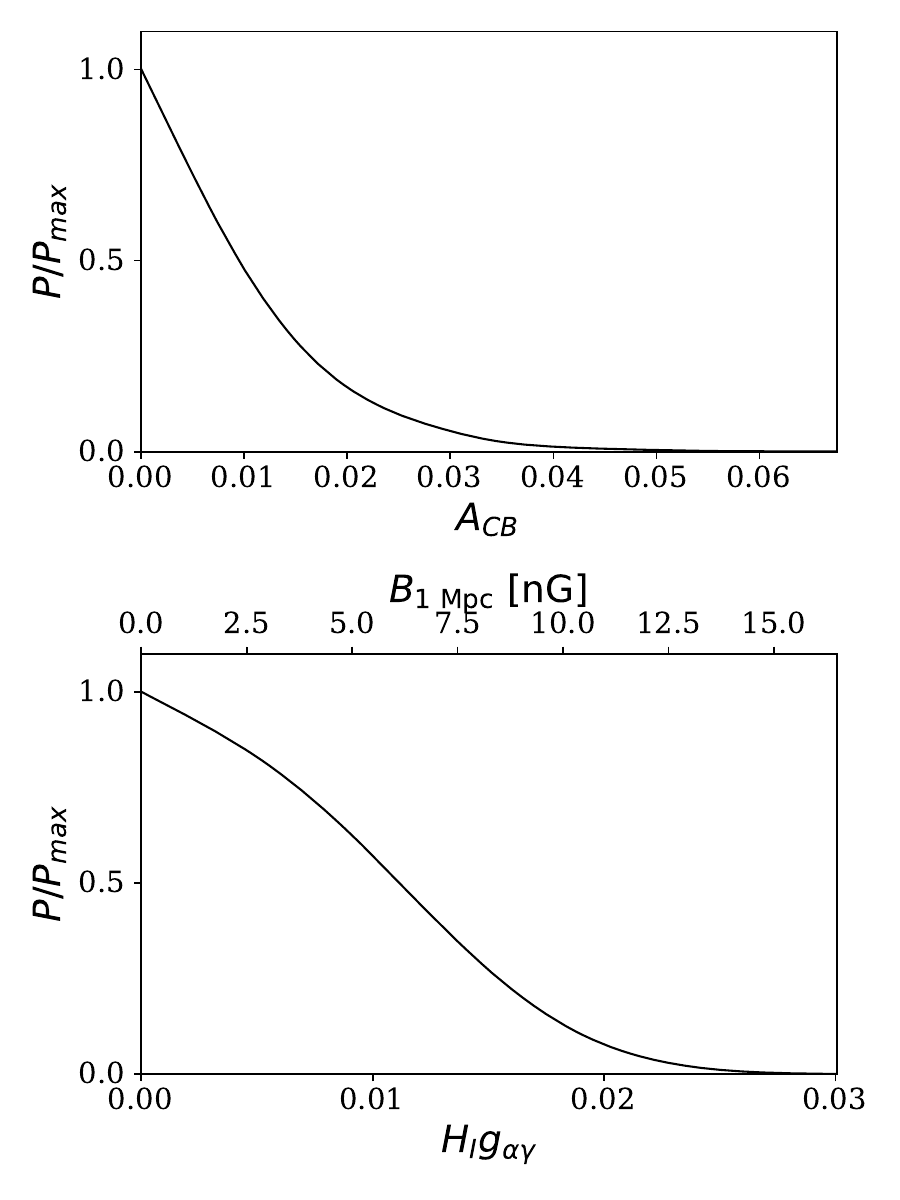}
    \caption{Top: Posteriors on \acb{}, defined by $\frac{L(L+1)}{2\pi} C^{\alpha\alpha}_{L} = A_{\textrm{CB}} \times 10^{-4}\ {\rm [rad^2]}$.
    The data show no preference for cosmic birefringence. Bottom: 1D posterior on product of the axion coupling constant $g_{\alpha\gamma}$ and Hubble rate at inflation $H_I$ and on \bpc{} derived from the weighted  bandpowers.
    }
    \label{fig:posteriors}
\end{center}
\end{figure}

We compare the measured rotation angle power spectrum to the scale-invariant power spectrum discussed in \S\ref{subsec:scaleinvar}. 
The posterior for \acb{} is shown in Fig.~\ref{fig:posteriors}; the data show no preference for cosmic birefringence. 
At 95\% confidence, we find the data set an upper limit of \upperscaleinvar{}. 
This limit improves upon the previous tightest constraint on $A_{\textrm{CB}}$ set by the BICEP/\textit{Keck} experiment of $A_{\textrm{CB}} < 0.044$ \cite{bicepkeck_XVII}. The BICEP/\textit{Keck} constraint is driven by the lowest-$\ell$ bin, which offers significant constraining power on a scale-invariant model despite the larger error bars.

\begin{figure}
\begin{center}
    \includegraphics[width=\linewidth]{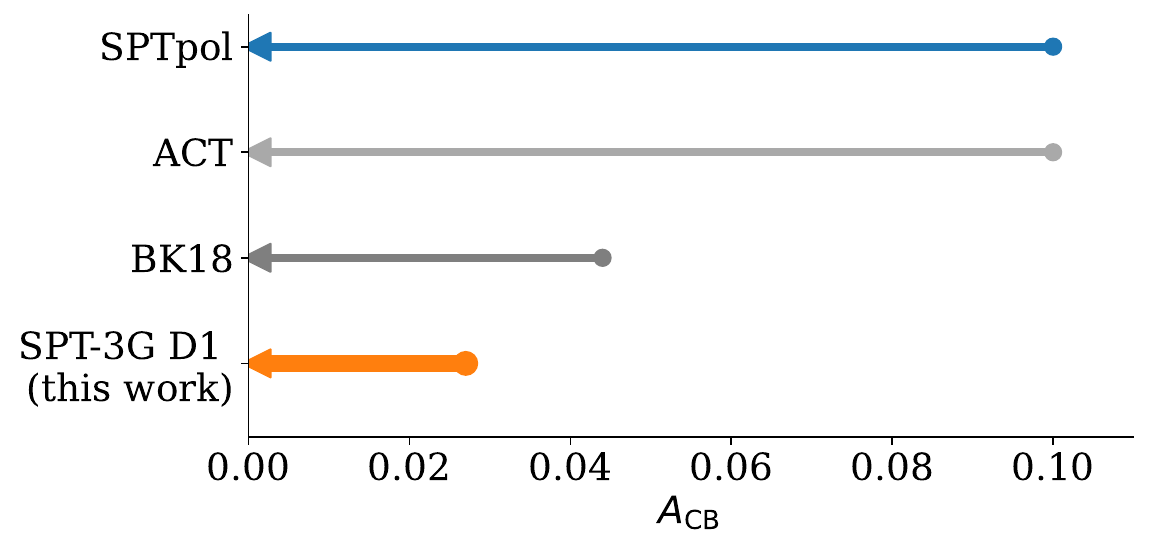}
    \caption{Comparison of the 95\% limits on $A_{\textrm{CB}}$ from SPT-3G (this work), BK18 \citep{ade23}, SPTpol \citep{bianchini20b}, and ACT \citep{namikawa20}.  
    }
    \label{fig:limitcompACB}
\end{center}
\end{figure}

This result is a factor of several improvement over the previous SPTpol search for anisotropic birefringence \citep{bianchini20b}, which set a limit of \upperscaleinvarsptpol{}. 
The tighter limit with SPT-3G compared to SPTpol is driven by reduction in noise levels, increase in sky area, the analysis change to the weighting used when binning the power spectrum, and to a lesser extent, the use of the exact curved-sky formalism. 
When uniform weights are used to bin, the SPT-3G upper limit weakens to  \upperscale{}. The observed improvement closely matches the expected improvement calculated from simulations.
We compare the limit on $A_{\textrm{CB}}$ from this analysis with other limits in the literature from the ACT \citep{namikawa20}, SPTpol \citep{bianchini20b}, and BICEP/\textit{Keck} \citep{ade23} CMB experiments in Fig.~\ref{fig:limitcompACB}. 
The polarization power spectrum can also be used to constrain anisotropic birefringence \citep{lonappan2025, Balkenhol_2025}, but at current experimental noise levels the 4-point estimator method used in this work yields much more stringent constraints. \citet{Balkenhol_2025} reports the tightest constraint using the polarization power spectrum to date, finding \upperscaleBB{} from SPT-3G measurements of the BB power spectrum.

This new limit on cosmic birefringence can be expressed as a limit on the product of the Hubble rate during inflation $H_I$ and axion coupling constant $g_{\alpha\gamma}$, using Eqn.~\ref{eqn:axion}. 
For a uniform prior on this product, the derived 95\% confidence limit on the axion coupling constant is \upperscaleginvar. 
The posterior on the product is shown in Fig.~\ref{fig:posteriors}. 
The new limit is the best current limit on $H_I g_{\alpha\gamma}$ from CMB polarization, improving over the previous best set by the BICEP/\textit{\textit{Keck}} experiment of $g_{\alpha\gamma} < 2.5 \times 10^{-2} /H_I$.

We use Eqn.~\ref{eqn:pmf} to translate the posterior for \acb{} to a posterior for \bpc{}, also resampling to apply a uniform prior in \bpc{}. 
We show the posterior for \bpc{} in Fig.~\ref{fig:posteriors}. 
The  upper limit on primordial magnetic fields from these 150\,GHz SPT-3G data is \upperscaleBinvar{} at 95\% confidence. 
Due to the steep frequency dependence of Faraday rotation, the 95\,GHz BICEP/\textit{Keck} data yields a comparable limit on primordial magnetic fields of $B_{\rm 1\,MPc} < 6.6$ nG at 95\% confidence \citep{bicepkeck_XVII}. 
Both searches find no evidence for primordial magnetic fields. 

The best current constraints on primordial magnetic fields at these scales come from two works. 
\citet{zucca17} analyze a combination of SPT and Planck polarization power spectra to find $B_{\rm 1\,Mpc} < 1.2$ nG.
\citet{paoletti22} use the effect PMFs would have on the universe's ionization history to constrain $\sqrt{\langle B^2\rangle} < 0.69$ nG. We expect significant improvement to our PMF constraints with the addition of the SPT-3G 95 GHz data.

We additionally search for birefringence from a collapsing string-wall network using the general white noise model discussed in \S\ref{subsec:scaledep} and find no evidence for birefringence. We place an upper limit of $A_{\textrm{wn}} < 0.023$ for the unweighted bandpower case. These results are consistent with \citet{jain2022} which found no evidence of cosmic birefringence from a collapsing string-wall network using \textit{Planck} (2018) and SPTpol data.

\section{\label{sec:conclusion}Conclusions}

We have presented the first constraints on anisotropic cosmic birefringence from the SPT-3G survey, reporting the measured power spectrum of the rotation angle in six bins over angular multipoles $20 < L < 2048$.
These data allow new and improved upper limits on the amplitude  of the scale-invariant rotation angle power spectrum: \upperscaleinvar{} at 95\% confidence. 

This limit on the rotation angle power spectrum can be recast into limits on either the axion coupling constant (which would rotate polarization through the Chern-Simons term) or the RMS strength of magnetic fields at recombination (which would rotate polarization through the Faraday effect).  
If all potential rotation is due to axions, we set a 95\% confidence upper limit on the axion coupling constant of \upperscaleginvar.

Instead assuming that the Faraday effect is the only source of possible rotation,  we set a 95\% confidence upper limit on the primordial magnetic field strength at 1\,Mpc scales of \upperscaleBinvar{}. 
This is comparable to the limit derived by \cite{ade23} from a similar analysis using 95\,GHz BICEP/\textit{Keck} data, but less constraining than upper limits from the universe's ionization history which yield $\sqrt{\langle B^2\rangle} < 0.69$ nG \citep{paoletti22}.

We additionally search for a white noise power spectrum sourced by anisotropic cosmic birefringence from collapsing axion string-wall networks. We find no evidence for this signal and place an upper limit on the white noise amplitude of $A_{\textrm{wn}} < 0.023$. As the theoretical landscape of axion networks continues to develop, CMB birefringence measurements will critically constrain these exotic models.

These results use two years of the SPT-3G survey data. 
The full survey will add $7$ more years of data to the same region of sky with an expected proportional $\sim 2.5\times$ reduction in the noise levels. Further improvements can be made by exploiting beyond-QE delensing methods to maximally isolate the birefringence signal \citep{yadav09, carron25, lonappan25}.
The SPT-3G 95\,GHz data has similar noise levels to the 150
\,GHz data, and will be invaluable for future searches for primordial magnetic fields due to the frequency dependence of the Faraday effect.  
We can expect even larger improvements when the next camera on the South Pole Telescope, SPT-3G+, begins taking data in 2029 \citep{natoli2026}.  
\\
\begin{acknowledgments}

The authors thank Andrew Long for the constructive discussions that guided this paper. 
The South Pole Telescope program is supported by the National Science Foundation (NSF) through awards OPP-1852617 and OPP-2332483. Partial support is also provided by the Kavli Institute of Cosmological Physics at the University of Chicago. This research was supported by The University of Melbourne’s Research Computing Services and the Petascale Campus Initiative.
Argonne National Laboratory’s work was supported by the U.S. Department of Energy, Office of High Energy Physics, under contract DE-AC02-06CH11357. 
The UC Davis group acknowledges support from Michael and Ester Vaida. 
Work at the Fermi National Accelerator Laboratory (Fermilab), a U.S. Department of Energy, Office of Science, Office of High Energy Physics HEP User Facility, is managed by Fermi Forward Discovery Group, LLC, acting under Contract No. 89243024CSC000002.
The Melbourne authors acknowledge support from the Australian Research Council’s Discovery Project scheme (No. DP260100705). 
The Paris group has received funding from the European Research Council (ERC) under the European Union’s Horizon 2020 research and innovation program (grant agreement No 101001897), and funding from the Centre National d’Etudes Spatiales. 
The SLAC group is supported in part by the Department of Energy at SLAC National Accelerator Laboratory, under contract DE-AC02-76SF00515.
Some of the results in this paper have been derived using the \texttt{healpy} \citep{zonca19} and \texttt{HEALPix} \citep{gorski05} package (https://healpix.sourceforge.net). Claude AI (Anthropic) was used to review code. The authors verified any changes.
\end{acknowledgments}



\bibliographystyle{apsrev4-2}
\bibliography{spt,mybib}

\end{document}